\documentclass[aps,prb,twocolumn,showpacs]{revtex4}

\def\cm{cm$^{-1}$}        % included by M. Hoevel on 2008-08-13
\usepackage{amsmath}    % included by M. Hoevel on 2008-07-16
\usepackage{graphics}% Include figure files
\usepackage{dcolumn}% Align table columns on decimal point
\usepackage{bm}% bold math

\begin{document}
\title{Dielectric properties of ultrathin metal films around the percolation threshold}
\author{Martin H{\"o}vel}
\author{Martin Alws}
\author{Bruno Gompf}
\email[]{gompf@pi1.physik.uni-stuttgart.de}
\author{Martin Dressel}
\affiliation{1.~Physikalisches Institut, Universit\"at Stuttgart, Pfaffenwaldring 57, D-–70550 Stuttgart, Germany}
\date{\today}

\begin{abstract}
We report on optical reflection measurements of thin Au films at
and around the percolation threshold (film  thickness 3 to 10~nm)
in an extremely broad spectral range  from 500 to 35\,000~\cm\
(0.3 -- 20~$\mu$m). Combining spectroscopic ellipsometry and
Fourier-transform infrared spectroscopy, the dielectric properties
of the films can be described over the whole frequency range by
Kramers-Kronig consistent effective dielectric functions. The
optical conductivity of the films is dominated by two
contributions: by a Drude-component starting at the percolation
threshold in the low frequency range and a plasmon in the
near-infrared region, which shifts down in frequency with
increasing film thickness. The interplay of both components leads
to a dielectric anomaly in the infrared region with a maximum of
the dielectric constant at the insulator-to-metal transition.

\end{abstract}


\maketitle

\section{Introduction}
Although for many applications closed metal films as thin as
possible are desired, the dielectric properties of percolating
metal films around the insulator-to-metal transition are not well
understood. Thick continuous metal films show a behavior similar
to bulk material \cite{Bennett66a} and can therefore be well
described by the Drude-model when corrections for size effects are
considered.\cite{Fahsold00a} With decreasing thickness, the films
become granular, and below the percolation threshold the metallic
behavior disappears as far as the electrical transport is
concerned. In principle one can try to simulate semi-continues
metal films with effective medium (EMA) theories
\cite{Maxwell1904a,Bruggeman35a}. However it was shown that EMA models fail to
describe the dielectric properties of ultra-thin metal films
\cite{Yagil92a, Gompf2007}. In general this transition region was not
yet studied in detail over a broad frequency range; nevertheless,
these investigations should eventually provide the data to link
the abrupt change in dc-conductivity with the observed shift of
plasmon resonances in the visible.

On the low frequency side of the electromagnetic spectrum,
percolation theories deal with an ``idealized'' dc-conductivity or
the electrical behavior at audio and radio frequencies up to some
MHz.\cite{Clerc90a} On the other side of the spectrum -- in the
visible and ultraviolet -- a large number of investigations have
been carried out to understand the optical properties of arrays of
clusters at surfaces.\cite{Bedeaux2002} In between, however -- in
the infrared and far-infrared spectral region -- little has been
done so far. Here we report on Fourier-transform infrared
reflection spectroscopy and spectroscopic ellipsometry
measurements of thin Au films on Si/SiO$_2$ covering the infrared
to ultraviolet spectral range between 500 to 35\,000~\cm\ (60~meV
to 4.3~eV), corresponding to a wavelength of 280~nm to 20~$\mu$m.
The film thickness $d$ was varied between 3 and 10~nm, i.\,e.\
from well below to well above the insulator-to-metal transition.

\section{Experimental Details}
Thin Au films were prepared by an electron-beam heated effusion
cell on clean $0.55$~mm thick Si(100)-substrates covered by a 200~nm
SiO$_2$ layer. The substrates were polished on both sides and held at room temperature during evaporation.
The preparation was performed in ultra high vacuum (UHV) at a base pressure of $1\cdot 10^{-7}$~Pa.
The deposition rate was about
$1$~\AA/min and measured prior to each film-deposition by a quartz crystal microbalance.
Keeping the gold-flux constant the film thickness could be controlled by the evaporation-time.
The film thicknesses vary from $d=3.4$~nm to $9.5$~nm.

All optical experiments were performed at room temperature. The
reflectivity measurements in the infrared (IR) were carried out by
a Bruker IFS 66/s Fourier transform infrared (FTIR) spectrometer
in the range $500$\,--\,$12\,000$~\cm\ employing a nitrogen cooled
mercury-cadmium-telluride (MCT) detector. All spectra were
recorded with a resolution of $0.5$~\cm\ and 64 averaged scans. As
reference a thick gold mirror was used. To compare measured
spectra with model calculations, etalon effects (multireflection
within the substrate) were smoothed out. For the ellipsometry
measurements a Woollam variable angle spectroscopic ellipsometer
(VASE) was utilized. The experiments were carried out in the
spectral range between $6000$ and $35\,000$~\cm\ (0.75 to 4.3~eV,
corresponding to a wavelength of 280~nm to 1.7~$\mu$m) with a
resolution of $170$~\cm\ and an angle of incident varied in steps
of $5^\circ$ from $35^\circ$ to $75^\circ$.\\

\section{Results and Analysis}
\begin{figure}
    \centering
    \scalebox{1.3}{\includegraphics{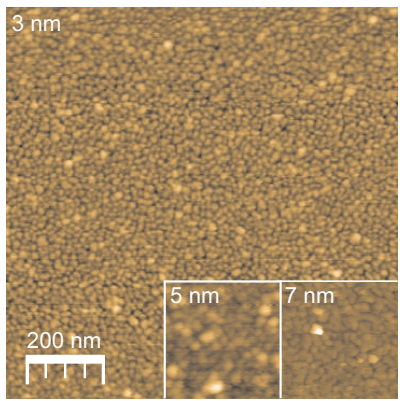}}
    \caption{(Color online) AFM images of 3~nm, 5~nm, and 7~nm thick gold films on Si/SiO$_2$.
    All three images are presented at the same scale. With increasing film thickness the cluster size increases and the films become smoother.}
    \label{fig:afm}
\end{figure}

The thickness dependent morphology of thin Au-films on Si/SiO$_2$
is well investigated \cite{Gompf2007, Pal04a}. In
Fig.~\ref{fig:afm} atomic force microscopy (AFM) images at the
same scale from a former investigation of 3~nm, 5~nm, and 7~nm
thick gold films are presented \cite{Gompf2007}.

With increasing film-thickness the size of the islands increases,
they get closer to each other and the films become smoother. The
root-mean-square roughness of the 3~nm film is about 1~nm, and the
average cluster size is about 4~nm. The percolation threshold
itself can not be determined from the morphology. Although it is
important to know the morphology of the films, one has to stress
the fact that percolation theories deal with scaling laws and
therefore the overall electrical and optical behavior should be
independent of the specific morphology \cite{Clerc90a}.

\begin{figure}
    \centering
    \scalebox{1.3}{\includegraphics{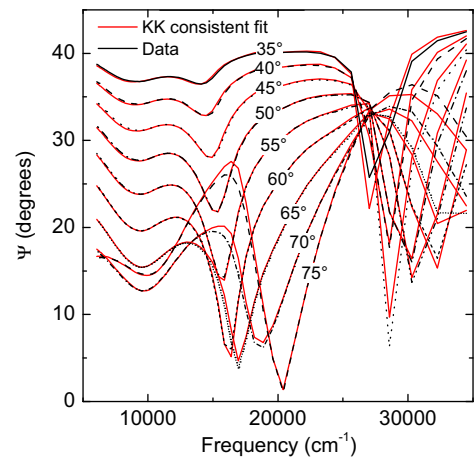}}
    \caption{(Color online) Measured ellipsometric angle $\Psi$ for a 8.1~nm thick Au film on a Si/SiO$_2$ substrate for several angles of incidence
    (dashed lines). The solid lines are the corresponding model fits. For details see text.}
    \label{fig:ellipsometry-psi}
\end{figure}

Prior to the film preparation, the bare Si/SiO$_2$ substrates were
characterized by ellipsometry and the obtained optical parameters
then kept constant in the further modeling of the films. As an example, in
Fig.~\ref{fig:ellipsometry-psi} the measured ellipsometric angle
$\Psi$ for a 8.1~nm thick Au film is shown as a function of frequency and angle of incidence. Several attempts were made
to model the obtained ellipsometric angles: It was not
possible, for instance, to model the granular films by an effective medium approximation (EMA), even in the case where the volume fraction was gradually decreased over the film thickness. Taking the
granular Au layer as a homogeneous effective layer with fixed
thickness, it was possible to get reasonable agreement to the
experimental data by a point-by-point fit with MSE-values around
6; but it turned out that the received dielectric functions were
not Kramers-Kronig consistent.

\begin{figure}
    \centering
    \scalebox{1.3}{\includegraphics{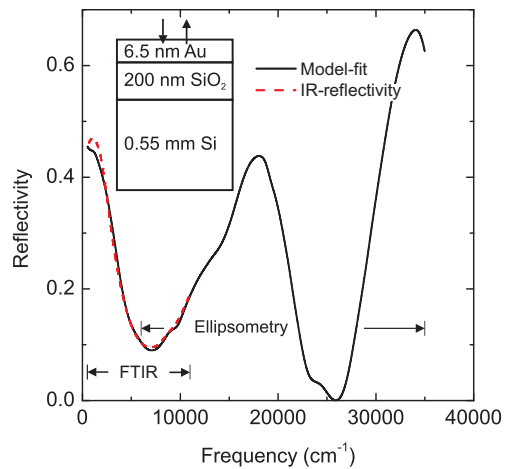}}
    \caption{(Color online) Measured IR-reflectivity of a $6.5$~nm Au film (dashed line) together with the reflectivity
calculated from the Kramers-Kronig consistent dielectric functions obtained by a simultaneous analysis of the IR- and ellipsometric data.
    The inset shows the assumed layer structure.}
    \label{fig:r}
\end{figure}

To solve this problem, we combined the results obtained by
ellipsometry and IR-reflectivity. Both sets of data were
simultaneously analyzed with the program package RefFIT. For
details on this tool see Ref.~\onlinecite{Kuzmenko2007}. By means
of a variational dielectric fit within the program, Kramers-Kronig
consistent dielectric functions could be obtained, which perfectly
reproduces both the ellipsometric angles (continuous line in
Fig.~\ref{fig:ellipsometry-psi}) as well as the IR-reflectivity.
In Fig.~\ref{fig:r}, for example, the measured reflectivity in the
IR-region (dashed line) of a $6.5$~nm thick Au film with its
corresponding fit (solid line) is shown together with an
extrapolation to the ultraviolet obtained from the ellipsometric
measurement.
The dominating oscillation in the reflectivity is due to
interferences caused by the 200~nm thick SiO$_2$ layer.

\begin{figure}
    \centering
    \scalebox{1.3}{\includegraphics{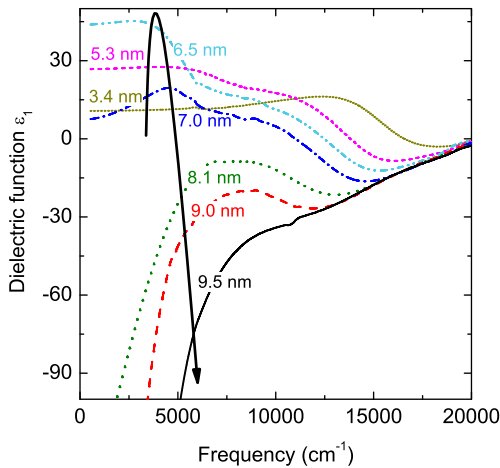}}
    \caption{(Color online) The effective dielectric constant of the gold films. Note the non-monotonous behavior at low frequencies as the film thickness $d$ increases (indicated by the arrow). From a gold-film thickness
of 8.1~nm on $\epsilon_1(\omega\rightarrow 0)$ is negative,
indicating the metallic behavior of the films.}
    \label{fig:e1-all}
\end{figure}

\begin{figure}
    \centering
    \scalebox{1.3}{\includegraphics{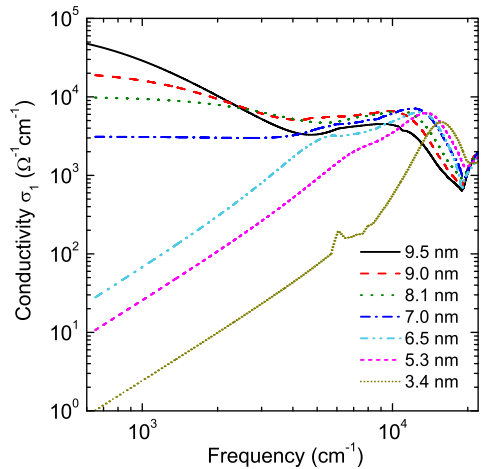}}
    \caption{(Color online) Optical conductivity of gold films of different thickness as indicated. The percolation occurs between 6.5 and 7.0~nm where the sign in the slope of $\sigma_1(\omega)$ changes at low frequencies.}
    \label{fig:s1-all-loglog}
\end{figure}

The real part $\epsilon_1(\omega)$ of the dielectric function as
received from the above analysis is shown in Fig.~\ref{fig:e1-all}
as a function of frequency for varying film thickness. From the
imaginary part $\epsilon_2(\omega)$ the optical
conductivity\cite{Dressel02a} was calculated by:
\begin{equation}
\sigma_1(\omega)=\omega\epsilon_0\epsilon_2(\omega),
\label{eq:drude}
\end{equation}
with $\epsilon_0$ the permittivity of free space. It is displayed
in Fig.~\ref{fig:s1-all-loglog} in a double logarithmic fashion.

The dielectric functions $\sigma_1(\omega)$ and $\epsilon_1(\omega)$ are described by a superposition of Drude-
and Lorentz-terms.\cite{Dressel02a} The parameters obtained from the fits for the different film thicknesses are listed in Table~\ref{tab:drude_lorentz}.

\section{Discussion}
\subsection{Percolation threshold}
\begin{table*}[tb]
\caption{The Drude- and Lorentz-parameters obtained from the fit to
the model. The plasma frequency is denoted by $\omega_p$ and
$\gamma=1/(2\pi c \tau)$ is the scattering rate with $c$ the speed
of light and $\tau$ the relaxation time. The bulk values are taken
from Ref.~\onlinecite{Bennett66a}.}\label{tab:drude_lorentz}
\begin{ruledtabular}
\begin{tabular}{ccccccccc}
film thickness&\multicolumn{2}{c}{Drude}     &     \multicolumn{3}{c}{1st Lorentz-oscillator}    &     \multicolumn{3}{c}{2nd Lorentz-oscillator}\\
$d$           &$\omega_p/2\pi c$ &$\gamma$   &$\omega_0/2\pi c$ &$\omega_p/2\pi c$ &$\gamma$     &$\omega_0/2\pi c$   &$\omega_p/2\pi c$ &$\gamma$\\
(nm)          &($10^4$ \cm)      &(\cm)      &(\cm)             &($10^4$ \cm)      &(\cm)        &(\cm)               &($10^4$ \cm)      &(\cm)  \\
\hline
bulk          &7.27              & 216       &--                &--                &--           &--                  &--                &--  \\
9.5           &5.98              & 795       &7218              &2.96              &5210         &10\,480             &2.76              &4960\\
9.0           &4.70              &1710       &6540              &3.60              &5920         &10\,400             &3.57              &4840\\
8.1           &4.79              &3770       &8280              &3.27              &5290         &11\,500             &3.18              &4230\\
7.0           &3.23              &5410       &7520              &3.71              &7090         &12\,220             &4.01              &5010\\
6.5           &--                &--         &6940              &3.86              &8280         &12\,710             &4.23              &5740\\
5.3           &--                &--         &9040              &3.27              &8540         &13\,620             &4.08              &5370\\
4.5           &--                &--         &--                &--                &--           &15\,650             &3.99              &5340\\
3.4           &--                &--         &--                &--                &--           &15\,600             &4.01              &5610\\
\end{tabular}
\end{ruledtabular}
\end{table*}
%film          Drude            Drude         Lorentz1       Lorentz1         Lorentz1           Lorentz2       Lorentz2         Lorentz2
%tickness      nu_p             gamma         nu_0           nu_p             gamma              nu_0           nu_p             gamma
%[nm]          [1/cm]           [1/cm]        [1/cm]         [1/cm]           [1/cm]             [1/cm]         [1/cm]           [1/cm]
%
%bulk          72718.727         215.808
%9.5           59790.97654       795.19073    7218.06352     29615.07938     5210.40458          10481.25831    27604.88534    4956.32584
%9.0           47028.43277      1706.42294    6536.32839     36027.52701     5916.61322          10397.27252    35656.8411     4839.53713
%8.1           47904.29176      3772.97298    8284.07668     32740.60457     5286.79524          11496.50752    31815.30594    4232.91548
%7.0           32294.88794      5407.78914    7524.12006     37068.47457     7094.1144           12218.282      40100.59629    5014.47896
%6.5           --               --            6935.27426     38615.7132      8283.2667           12711.90878    42277.2742     5739.6241
%5.3           --               --            9039.29724     32656.39564     8542.15049          13616.01843    40846.62986    5367.63967
%4.5           --               --            --             --              --                  15646.53307    39947.65511    5342.22975
%3.4           --               --            --             --              --                  15598.70425    40078.99111    5608.76952
%
From the effective conductivity $\sigma_1(\omega)$ as well as from
the effective dielectric function $\epsilon_1(\omega)$ of the gold
layers, the different films can be clearly divided in two regimes:
The continuous films in the thickness range $d\geq7$~nm and the
granular films with $d\leq6.0$~nm. The percolation threshold $d_c$
falls right between these two values.

Well above the percolation threshold, the conductivity decreases with frequency and the static permittivity $\epsilon_1(\omega\rightarrow 0)$ is negative, both
indicating a typical metallic behavior.
The low-frequency behavior of the metallic films can be fitted by a simple Drude model\cite{Dressel02a} with
plasma frequencies $\omega_p$ and scattering rates
$\gamma=1/(2\pi c\tau)$ as summarized in Table~\ref{tab:drude_lorentz}.
In addition, the fit parameters of two Lorentz oscillators
(discussed below) are listed. Naturally, the reflectivity of the evaporated
films is lower compared to bulk Au, because the effective
electron density is reduced and the scattering rate $\gamma$ becomes larger with decreasing film thickness.
While the former effect comes from a surface dipole layer\cite{Lang1973, Schulte1976}, the latter is a
well-known consequences of the classical size effect. Both effects
become more pronounced as $d$ is reduced since the
surface-to-volume fraction is enhanced.\cite{Pucci06a, Walther07a, Brandt2008}

Below the metal-to-insulator transition (MIT) the static
conductivity $\sigma_1(\omega\rightarrow 0)$ vanishes and
$\epsilon_1(\omega)$ is positive. The optical behavior in this
regime can be explained by the absence of the Drude contribution
and two resonance peaks in the mid-infrared. Below the MIT we
picture metallic islands which interact capacitively. While the
dc-conductivity is zero, the optical conductivity depends on the
capacitive coupling of the islands and thus increases with
frequency. From the AFM images of comparable samples (see Fig.~\ref{fig:afm})
and other very extensive studies\cite{Pal04a} of the morphology of Au on Si/SiO$_2$,
the thickness dependence can be explained as follows:
the metallic clusters are separated from each other forming a condenser.
With increasing effective film thickness $d$ the islands are closer to each other and also larger.
While their surface increases, the spacing decreases and tends to zero at the percolation
threshold. This leads to a rise of the capacitive coupling and
an enhanced optical conductivity (Fig.~\ref{fig:s1-all-loglog}). For a detailed
discussion from a more theoretical point of view see e.\,g.\ Ref.~\onlinecite{Dubrov1976, Efros76a}.

\subsection{Dielectric constant}
As far as the thickness dependence of the low-frequency permittivity is concerned, one would expect that $\epsilon_1$, which is positive for $d=0$,
gradually decreases with $d$ and becomes negative at the MIT.
However, a distinctively different behavior is observed in our experiments.
As indicated by the arrow in Fig.~\ref{fig:e1-all},
$\epsilon_1(\omega)$ first rises with film thickness $d$,
goes through a maximum at some critical concentration $d_c$ which is assumed to be around 6.0~nm,
and then decreases rapidly; $\epsilon_1$ becomes negative
only at considerably larger $d$, above 7~nm.

In principle, this behavior is known from the low-frequency
conductivity of percolating networks. Efros and Shklovskii
investigated such systems theoretically and predicted a
divergence of the static dielectric constant
\begin{equation}
\epsilon_1(0,d)\propto(d_c-d)^{-s} \label{eq:div}
\end{equation}
at a critical thickness $d_c$.\cite{Efros76a}
For any small but finite frequency
$\omega$, the divergence simply becomes a maximum:
\begin{equation}
\epsilon_1(\omega,
d_c)=\epsilon_s\left(\frac{\sigma_m}{\omega\epsilon_0\epsilon_s}\right)^{1-u}\quad ,
\label{eq:max}
\end{equation}
where $\sigma_m$ is the real part of the conductivity of the
metallic fraction and $\epsilon_s$ the real part of dielectric
constant of the substrate. For a two-dimensional system the
critical exponents in Eqs.~(\ref{eq:div}) and (\ref{eq:max}) are
$s=1.3$ and $u=0.5$. For three dimensions $s=1$ and
$u=0.62$.\cite{Efros76a}
\begin{figure}
    \centering
    \scalebox{1.3}{\includegraphics{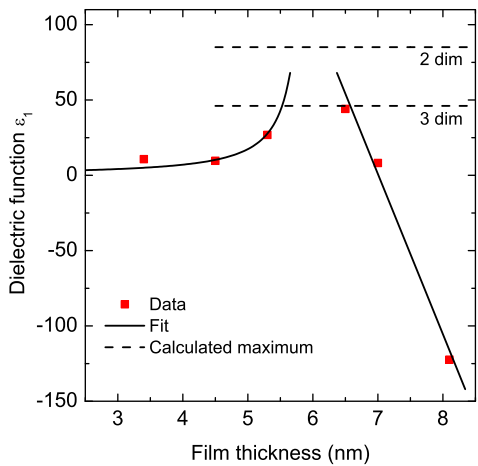}}
    \caption{(Color online) Divergence of the dielectric constant at 1000~\cm. The solid line is a fit to the data: $\epsilon_1\propto(6-d)^{-1.3}$ for $d<6$~nm and linear above. The dashed lines represent the predicted maximum after Eq.~(\ref{eq:max}) for two and three dimensions.}%epsilon_1=17.38865*(6-x)^(-1.3)
    \label{fig:divergence-e1}
\end{figure}
Figure~\ref{fig:divergence-e1} exhibits the dielectric constant as
a function of film thickness measured at a frequency of 1000~\cm;
it represents more or less the static value since $\epsilon_1$ is
almost frequency independent below this value. The measurements
fit to the expected behavior with a maximum at the critical
thickness $d_c$ somewhere around 6~nm. The zero-crossing of
$\epsilon_1(\omega=0,d)$ occurs at $d_0$ slightly above 7~nm,
where the Drude-component in the films start to develop (cf.\
Table~\ref{tab:drude_lorentz}). The solid line corresponds to
Eq.~(\ref{eq:div}) with $d_c=6$~nm and $s=1.3$ but does not
unambiguously refer to a two-dimensional system since the data can
also be explained with the same critical thickness but the
three-dimensional critical exponent $s=1$ (not shown). The dashed
lines indicate the maximum value of the dielectric constant
$\epsilon_1(1000~{\rm cm}^{-1}, d_c)$ for two and three dimensions
according to Eq.~(\ref{eq:max}).

For a more accurate analysis directly at the percolation
threshold and a quantitative prove of equation (\ref{eq:max}),
additional data with closer steps in thickness are required.
It is interesting to note that the 7~nm film still exhibits a positive $\epsilon_1$ (cf.\ Fig.~\ref{fig:e1-all}) albeit it possesses a Drude component.

\subsection{Plasmons}
\begin{figure}
    \centering
    \scalebox{1.3}{\includegraphics{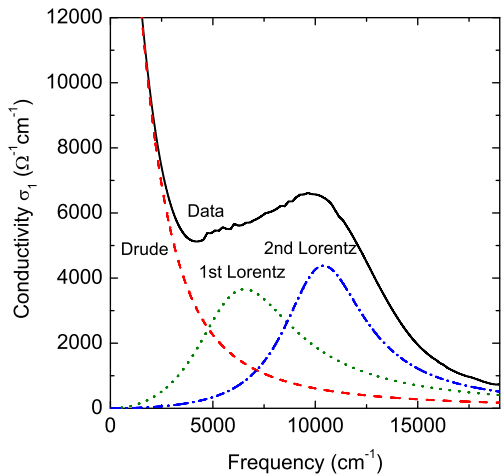}}
    \caption{(Color online) Decomposition of the measured conductivity of the 9~nm film in its Drude-
    and Lorentz-components as listed in Table~\ref{tab:drude_lorentz}.}
    \label{fig:conductivity-components}
\end{figure}

For the description of the behavior at higher frequencies
additional Lorentz-oscillators have to be considered. To illustrate
how the conductivity is built up by the different components,
the conductivity of the 9~nm film is shown in
Fig.~\ref{fig:conductivity-components} together with its different contributions. With
increasing film thickness the two oscillators shift to
lower frequencies, become weaker and narrower (cf.~Table~\ref{tab:drude_lorentz}).

In the following we will first concentrate on the second
oscillator at higher frequencies which is caused by the ensemble
of single particles. It is a property of the ensemble and has to
be distinguished from free electron oscillations in single metal
clusters. The incident electric field is modified by the
polarizability of these particles which -- most important --
interact.\cite{Doremus66a,Marton71a} In literature this
transverse-mode plasmon is referred to as ``Maxwell-Garnett
resonance'' or ``optical conduction resonance''. It shifts to
lower frequencies upon stronger interaction as the metal-islands
become closer and denser packed. Assuming spherical clusters with
a diameter much smaller than the used  infrared wavelength, the
position of this plasma absorption can be described by the
following condition~\cite{Doremus66a}:
\begin{equation}
\epsilon_{1}^{\rm bulk}(\omega)= \epsilon_{1} = -\frac{2+Q}{1-Q}n_s^2\label{eq:plasmon_condition}
\quad .
\end{equation}
In this condition, $Q$ is the area fraction of the substrate
covered by metal, $n_s$ is the refractive index of the substrate
and $\epsilon_{1}^{\rm bulk}(\omega)$ is the frequency dependent
dielectric constant of the metallic fraction. For a given $Q$, we can
calculate the right-hand side of the equation and obtain a certain
dielectric constant $\epsilon_1$. The frequency at which the obtained
resonance condition $\epsilon_{1}^{\rm bulk}(\omega)=\epsilon_1$
is fulfilled gives the position at which the plasmon appears.

\begin{figure}
    \centering
    \scalebox{1.3}{\includegraphics{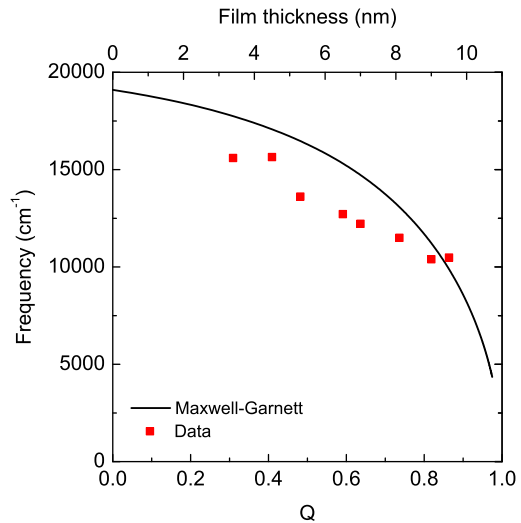}}
    \caption{(Color online) The position of the plasmon shifts to lower frequencies with increasing film thickness.
 The solid line is calculated from Eq.~(\ref{eq:plasmon_condition}) with the values of bulk gold\cite{Hovel} and belongs to the $Q$-axis.
 The dots show the experimental position of the plasmon for films with thicknesses as given by the upper-axis.
 The proportional relation between the two abscissas is arranged in the way that the slope of the experimental and calculated values fits best.}
    \label{fig:plasmon}
\end{figure}

If we use the frequency dependent values
$\epsilon_{1}^{\rm bulk}(\omega)$ of bulk gold as calculated
from \cite{Hovel} and set $n_s=1.45$ for the
SiO$_2$-substrate, we get the solid line in
Fig.~\ref{fig:plasmon}. We can now compare this line with the
measured plasmon frequencies in dependence of film thickness as
listed in Table~\ref{tab:drude_lorentz} as second Lorentz-oscillator $\omega_0$.
Here we assume a linear relation between the film thickness
$d$ and the area fraction $Q$ covered by the metal.
As seen from Fig.~\ref{fig:plasmon}, the
plasmon shifts to lower frequencies with increasing film thickness $d$,
nicely following the behavior given by Eq.~(\ref{eq:plasmon_condition}).

A closer inspection yields small deviations of the measured
plasmon frequencies to lower values, which we ascribe to the influence of
the substrate.
The optical properties of the 200~nm oxide layer might be influenced by the
underlaying silicon, which has a much higher refractive index.
Taking a somewhat larger $n_s$ in Eq.~(\ref{eq:plasmon_condition})
shifts the calculated values
to lower frequencies, i.\,e.\ closer to the experimental data.

As can be seen from Fig.~\ref{fig:plasmon} for $Q\rightarrow1$ the
calculated plasmon frequency shifts to zero and thereby mutates to
a Drude-peak. It is known that the Maxwell-Garnett theory breaks down
above the percolation threshold, which in our case is at about
$Q=0.6$, therefore the extrapolation to $Q\rightarrow 1$ is
disputable. Nevertheless, the experimentally found position of this
plasmon can still be described quite well within this theory even
above the percolation threshold (see Fig~\ref{fig:plasmon}).

Recently de Vries {\em et al.}\cite{Vries07a} interpreted their ellipsometric
measurements on thin silver films in the visible and near-infrared spectrum
exactly in this way: the resonance frequency of a localized plasmon
shifts to lower frequencies with increasing film thickness becoming
zero at the percolation threshold. There the relaxation
time exhibits an abrupt increase indicating the transition to a
macroscopic conducting state.

Our measurements, which extent to the far infrared, reveal a
different picture. At the percolation threshold we can clearly see
both the development of a Drude-peak and the presence of the
plasmon. Even well above the percolation threshold, the position
of this plasmon can still be described satisfactorily by
Eq.~(\ref{eq:plasmon_condition}), but it gradually dies out with
increasing film thickness $d$. This can in principle be
interpreted in the way that the rough surface of the percolated
films still show dipole interaction or that dielectric inclusions
in the film start to interact with each other, as it has been
considered by Cohen \textit{et al.}\cite{Cohen73a}. In a direct
comparison of the conductivity of the calculated and measured
plasmon the Lorentzian-line of the latter one is broadened (data
are not shown). This comes from the fact that the particle size is
smaller than the mean free path of the conduction electrons. As
seen before, this size effect also leads to a large scattering
rate in the Drude component of the percolated films (cf.\
Table~\ref{tab:drude_lorentz}). Therefore the Maxwell-Garnett
theory gives correctly the resonance position but the calculated
plasmon band is wider than expected. The occurrence of this effect
for thin films was already predicted by Doremus \cite{Doremus66a}.

As discussed above the plasmon referred to as second Lorentz
oscillator in Table~\ref{tab:drude_lorentz} is a property of an
ensemble of single clusters. At low coverage only this peak is
seen. At higher coverage a second peak at lower frequencies
appears. This mode is an indication for the formation of
aggregates \cite{Kreibig1986}. With increasing film thickness both
plasmon peaks shift to lower frequencies, they broaden and the
splitting between them increases. In the classification given by
Kreibig \textit{et al.}\cite{Kreibig1986} the sample undergoes a transition form
category II (separated single clusters in full statistical
disorder) to category IV (various kinds of aggregates, various
next neighbor distances, plus single particles). At the
percolation threshold, when additionally the Drude component
appears, they can not be clearly separated anymore (see
Fig.~\ref{fig:conductivity-components}).

\begin{figure}
    \centering
    \scalebox{1.3}{\includegraphics{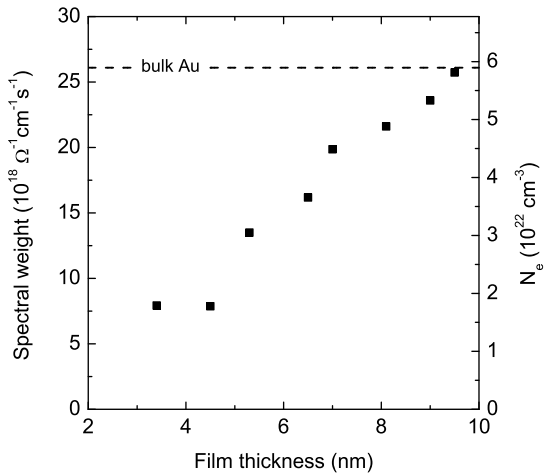}}
    \caption{Spectral weight and electron density $N_e$ of the Drude- and the two Lorentz-oscillators as function of the film thickness.
 The dashed line represents the electron density as determined for bulk gold.}
    \label{fig:spectral-weight}
\end{figure}

The maximum in $\epsilon_1(\omega)$ at the critical thickness at
low frequencies is now a direct consequence of the two competing
contributions: below the percolation threshold the shift of the
plasmons to lower frequencies with increasing film thickness leads
to an increasing $\epsilon_1(\omega)$. At the percolation
threshold $d_c$ the Drude-peak starts to develop, which adds a
strong negative component to the dielectric response leading to a
maximum in the dielectric function. Eventually $\epsilon_1$
changes sign at $d_0$ which is slightly above $d_c$. This
interpretation is supported by the analysis of the spectral weight
and the corresponding electron density $N_e$ shown in
Fig.~\ref{fig:spectral-weight}. The spectral weight increases
linearly with film thickness reaching the bulk value at about
10~nm. No indication of the MIT between 6.0~nm and 7~nm can be
identified, i.\,e.\ the abrupt change in dc-conductivity is not
reflected at higher frequencies. There a monotonous transfer of
spectral weight from the plasmons to the Drude-peak is observed.
The linear increase of the electron density simply describes the
growing amount of metal. A similar result for thin silver films
was observed by de Vries \textit{et al.}\cite{Vries07a}.

\section{Conclusion}
Combining FTIR spectroscopy and spectroscopic ellipsometry the
effective dielectric function of thin Au films around the
percolation threshold could be obtained over a very broad
frequency range from the infrared up to the UV. The optical
properties of the films can in principle be described by two
contributions: a plasmon in the near-infrared, which shifts down
with increasing film thickness and than slowly dies out above the
MIT, and a Drude peak, which starts to develop at the MIT and
than rapidly increases with film thickness. The
interplay of both components leads to a dielectric anomaly, known
from percolation theory: from dc up to a few thousand wavenumbers
$\epsilon_1(\omega)$ exhibits a pronounced maximum at
some critical thickness.

\end{document}